\documentclass[12pt]{article}
\usepackage{epsfig}
\newcommand{\eq}{\begin{equation}}
\newcommand{\eqx}{\end{equation}}
\newcommand{\eqn}{\begin{eqnarray}}
\newcommand{\eqnx}{\end{eqnarray}}
\newcommand{\dt}{\Delta}

\newcommand{\dd}{\partial}
\newcommand{\lab}{\label}
\begin{document}
%
\begin{center}
{\large\bf Polarised deep inelastic scattering accompanied by a forward jet
as a probe of the  $\ln^2(1/x)$ resummation}
\vspace{1.1cm}\\
{\sc J.~Kwieci\'nski,}\footnote{e-mail: jkwiecin@solaris.ifj.edu.pl}
{\sc B. Ziaja}\footnote{e-mail: beataz@solaris.ifj.edu.pl}\\
\vspace{0.3cm}
{\it Department of Theoretical Physics,\\
H.~Niewodnicza\'nski Institute of Nuclear Physics,
Cracow, Poland}
\end{center}
\vspace{1.1cm}
\begin{abstract}
We argue that the production of forward jets in polarised deep
inelastic scattering can be a useful tool for probing the double
$ln^2(1/x)$ resummation effects which control the
polarised deep inelastic scattering for small values of the\\
Bjorken parameter $x$.  We solve the corresponding integral
equations generating the double $ln^2(1/x)$ resummation and
calculate the differential structure function  describing the  
forward jet production in the
small $x$ regime which can possibly be probed by the polarised
HERA measurements.  We show that these structure functions
should exhibit the characteristic increase with decreasing $x/x_J$,
where $x_J$ denotes the
longitudinal momentum fraction of the parent proton carried by
a jet, and we quantify this increase. 
\end{abstract}
\vspace{1.1cm}
The idea of studying the deep inelastic $ep$ scattering with an identified
forward jet as a probe of the low $x$ behaviour of QCD was proposed
by Mueller \cite{MUELLER}. Since then it has been successfully applied to the
case of unpolarised deep inelastic scattering (DIS) \cite{FJET1,FJET2}. 
The Mueller's proposal was
to study the unpolarised deep inelastic events $(x,Q^2)$ containing
an identified forward jet $(x_J,k^2_{J})$, where 
 the longitudinal momentum fraction $x_J$ of the proton carried by
a forward jet and jet transverse momentum squared $k^2_{J}$ 
were assumed to fulfill the conditions~:
\eqn
x_J & \gg & x,\lab{assumx}\\
k^2_{J} & \sim & Q^2\lab{assumk}.
\eqnx
Here, as usual, $Q^2=-q^2$, where $q$ is the four momentum transfer between 
electrons in the deep inelastic $ep$ scattering, and $x$ is the Bjorken 
parameter, i.e. $x=Q^2/(2pq)$ with $p$ denoting the four momentum 
of the proton.

The first assumption $x_J\gg x$ implies that one probes the QCD dynamics
in the low $x/x_J$ region (the quantities which are measured depend
on the ratio $x/x_J$). The second assumption $k^2_{J} \sim Q^2$
guarantees  suppression
of the standard leading order (LO) DGLAP evolution \cite{MUELLER,FJET1,FJET2} and restricted
penetration of the non-perturbative region by the small $x$ resummation
\cite{CIGAR}.
Let us recall that the small $x$ behaviour of (unpolarised)
deep inelastic  scattering should be controlled  by the QCD
pomeron which in perturbative QCD is described  by the Balitskij,
Fadin, Kuraev, Lipatov (BFKL) equation \cite{BFKL}.  This equation generates
the leading $ln(1/x)$ resummation, and it
corresponds to the sum of ladder diagrams with the (reggeised)
gluons along the chain.  Unlike the LO DGLAP equations which
correspond to  ladder diagrams with ordered transverse momenta,
the  transverse momenta of the gluons along the BFKL ladder are not
ordered. In fact, the small $x$ behaviour generated by the BFKL
equation is linked with the diffusion of the transverse momenta both
towards the infrared and the ultraviolet regions. It is this
diffusion which is probed in the kinematical configurations which have
comparable scales on both sides of the ladder,  i.e. $k_J^2 \sim Q^2$
for the forward jet production.

In this paper we would like to apply the idea of forward jet measurement
to the case of polarised deep inelastic $ep$ scattering  in the region of small values of $x$ which can 
be probed at the possible polarised HERA measurements \cite{ALBERT}. 
The pomeron contribution decouples from the
polarised deep inelastic scattering, and its small $x$ limit is sensitive
to the novel effects of the double $ln^2(1/x)$ resummation \cite{BARTNS,BARTS}.

The relevant dynamical quantity in the case of  jet production 
in  polarised deep inelastic scattering  is the differential spin 
structure function $x_J{\partial g_1\over \partial x_J \partial
k_j^2}$. It is linked to the corresponding differential cross-section
in the standard way:
\eq
{\dd^4 \sigma \over \dd x \dd Q^2 \dd x_J \dd k_J^2}=
\frac{8\alpha_e^2\pi}{Q^4}\,\,\,
{\dd^2 g_1 \over \dd x_J \dd k_J^2}
\,\, y (2-y),
\lab{cross}
\eqx
where $\alpha_e$ denotes the electromagnetic coupling constant, and $y$ describes
the energy fraction of incoming electron carried by the interacting virtual photon.  
In equation (\ref{cross}) we have omitted terms proportional to $\gamma^2 = 
4M^2x^2/Q^2$, where $M$ denotes nucleon mass,  which are negligible 
at small $x$. 
The cross section in formula (\ref{cross}) corresponds, as usual, to the 
difference between the cross-sections for antiparallel and parallel spin 
orientations \cite{SMC}.  Similarly as for the unpolarised case, restrictions (\ref{assumx}),
(\ref{assumk}) applied to the polarised DIS forward jet events
allow one to neglect the effects of DGLAP evolution
and to concentrate on the low $x/x_J$ behaviour of
$x_J{\dd^2 g_1 \over \dd x_J \dd k_J^2}$.
Low $x/x_J$ behaviour of differential spin structure function is
dominated by double logarithmic  $ln^2(1/x)$ (or rather $ln^2(x_J/x)$)  
contributions
i.e. by those terms of the perturbative expansion which correspond to
the powers of $ln^2(x_J/x)$ at each order of the expansion, similarly
as for the DIS spin dependent structure function $g_1$ \cite{BARTNS,BARTS}.
In the following we will apply an approach to the double $ln^2(1/x)$
resummation based on unintegrated parton distributions
\cite{BBJK,MAN,BZIAJA, BZIAJA2}.
\begin{figure}[t]
    \centerline{
     \epsfig{figure=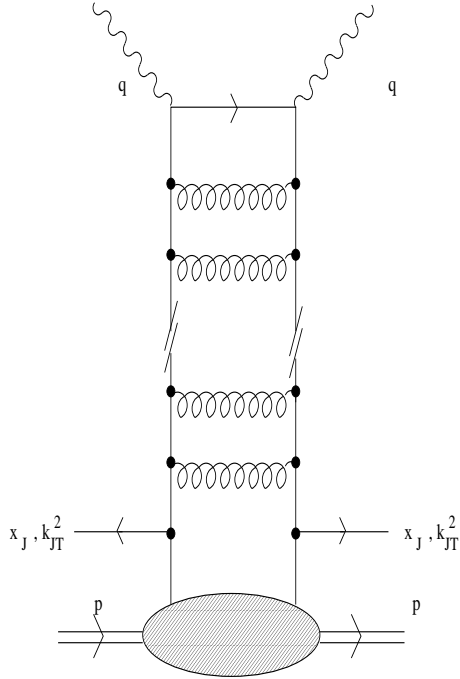,height=9cm,width=6cm}
               }
     \caption{An example of the ladder diagram contributing to the differential
spin structure function.}
\label{fig.1}
\end{figure}

The full contribution to the double $ln^2(1/x)$ resummation comes from
the ladder diagrams with quark and gluon exchanges along the ladder
(see e.\ g.\ Fig.\ 1) and the non-ladder bremsstrahlung diagrams \cite{QCD}.
The latter ones are obtained from the ladder diagrams by adding to them
soft bremsstrahlung gluons or soft quarks \cite{BARTNS,BARTS,QCD}, and they
generate the infrared corrections to the ladder contribution. However,
it was shown in ref. \cite{BZIAJA2} that the effects of non-ladder contributions
considered for DIS spin dependent structure function $g_1$ are visible only for 
$x < 10^{-3}$. In what follows we assume $x/x_J>10^{-3}$ and limit ourselves to the
contributions from ladder diagrams which dominate in this region.

The relevant region of phase space generating the double
$ln^2(1/x)$ resummation from  ladder diagrams corresponds
to ordered $k_n^2/x_n$, where $k_n^2$ and  $x_n$ denote respectively the
transverse momenta squared and longitudinal momentum fractions of the proton 
carried 
by  partons  exchanged along the ladder \cite{QED}.  In consequence, the effects 
of the double $ln^2(1/x)$ resummation should be still visible 
in the kinematical configuration  having comparable scales 
on both sides of the
ladder (i.e. $k_J^2 \sim Q^2$ for jet production)  in contrast to the LO DGLAP evolution alone which
corresponds to ordered transverse momenta.  

\begin{figure}[t]
    \centerline{
     \epsfig{figure=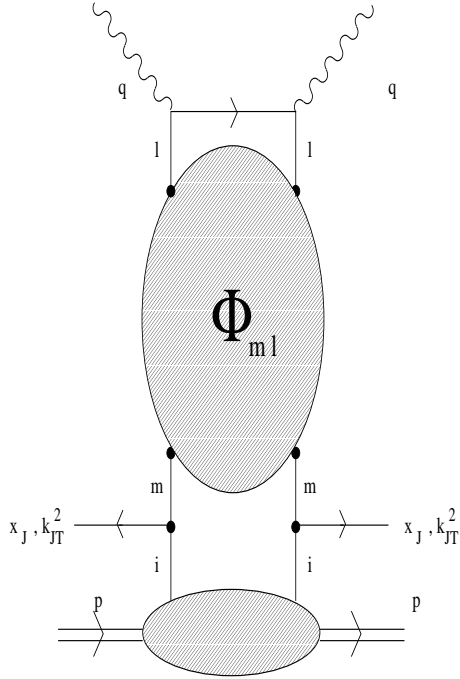,height=9cm,width=6cm}
               }
     \caption{Diagrammatic representation of formula (\ref{dg1}) for the 
     differential structure function $x_J{\dd^2 g_1\over \dd x_J\dd k_J^2}$ 
     describing the (forward) jet production in polarised deep inelastic 
     scattering.}
\label{fig.2}
\end{figure}

The formula for the differential structure function can be written in
the following form (see Fig.\ 2)~:
\begin{equation}
x_J{\dd^2 g_1\over \dd x_J\dd k_J^2}
= \bar \alpha_s(k_J^2)\sum_{iml}\Delta p_i(x_J,k_J^2)\Delta P_{mi}(0)
\Phi_{ml}({x\over x_J},k_J^2,Q^2),
\label{dg1}
\end{equation}
where $\Delta p_i(x_J,k_J^2)$ are the (integrated) spin dependent parton distributions
in the proton.  The quantities $\Delta P_{im}(0) \equiv \Delta P_{im}(z=0)$,
where $\Delta P_{im}(z)$ (for quarks $\Delta P_{im}(z)=\delta_{im}\dt P_{qq}$)
denote the LO splitting functions describing evolution
of spin dependent  parton densities. The indices $i$ and  $m$ numerate quarks, antiquarks
and gluons and the index $l$ quarks and antiquarks respectively.
Finally, $\bar \alpha_s = \alpha_s/2\pi$. Equation (\ref{dg1}) was derived,  
assuming  strong ordering of transverse momenta 
($k_i^2 << k_J^2 \sim k_m^2$) and of the longitudinal momentum fractions 
($x_m << x_i \sim x_J$) at the jet production vertex (see Fig.\ 2).  

\noindent
The functions $\Phi_{ml}({x\over x_J},k_J^2,Q^2)$ are related to the
unintegrated  quark and antiquark distributions
$f_{ml}({\bar x\over x^{\prime}}, k_J^2, k_f^2)$ in the parton $m$,
where $ k_f^2$ denotes the transverse momentum squared of the quark (antiquark):
\begin{equation}
\Phi_{ml}({x\over x_J},k_J^2,Q^2)={1\over 2} e_l^2
\int^{\bar W^2} {dk_f^2\over k_f^2}
\int_{\bar x}^{x_J} {dx^{\prime}\over x^{\prime}}
f_{ml}({\bar x\over x^{\prime}},k_J^2, k_f^2), 
\label{phiml}
\end{equation}
where
\eqn
\bar x &=&x\left(1 + {k_f^2\over Q^2} \right),\label{barx}\\
\bar W^2 &=& Q^2\left({x_J\over x} - 1 \right).\label{barw2}
\eqnx

We show later that the integration limits in eq. (\ref{phiml}) are further 
restricted by the ordering of $k_n^2/x_n$.  In particular this ordering will 
provide the infrared cut-off for $k_f^2$. \\

It is convenient to introduce the non-singlet and singlet combination of
parton distributions
\begin{equation}
f^{NS}_m = \sum_{l=1}^{N_f} \left({e_l^2\over\langle e^2 \rangle} 
- 1\right)f_{ml},
\label{fns}
\end{equation}
\begin{equation}
f^S_m = \sum_{l=1}^{N_f}f_{ml},
\label{fs}
\end{equation}
where $\langle e^2\rangle = {1\over N_f}\sum_{l=1}^{N_f}e_l^2$, and $N_f$ denotes the
number of active flavours ($N_f=3$).

\noindent
Restricting ourselves to ladder diagrams in  the double $ln^2(1/\xi)$ 
approximation,  
 we get the following equations for
the functions $f_m^{NS}(\xi,k_J^2, k_f^2)$ and
$f_m^{S}(\xi,k_J^2, k_f^2)$:
{\footnotesize
\eqn
\nonumber\\
& &f_m^{NS}(\xi,k_J^2, k_f^2) = f_m^{NS 0}(\xi,k_J^2, k_f^2)+\nonumber\\
& &\bar \alpha_s(\mu^2)
\Delta P_{qq}(0)\int_{\xi}^1 {d\xi^{\prime}\over \xi^{\prime}} 
\int_0^{k_f^2/z} {dk^2\over k^2}
f_m^{NS}(\xi^{\prime},k_J^2, k^2)\Theta(k^2-k_J^2 \xi^{\prime})
\label{ns},\\
\nonumber\\
& &f_m^{S}(\xi,k_J^2, k_f^2)=f_m^{S 0}(\xi,k_J^2, k_f^2)+\nonumber\\
& &\bar \alpha_s(\mu^2)
\int_{\xi}^1 {d\xi^{\prime}\over \xi^{\prime}} \int_0^{k_f^2/z} {dk^2\over k^2}
\left(\Delta P_{qq}(0)f_m^{S}(\xi^{\prime},k_J^2, k^2) + \Delta P_{qg}(0)
f_m^{g}(\xi^{\prime},k_J^2, k^2)\right)\Theta(k^2-k_J^2 \xi^{\prime})
\label{s},\\
\nonumber\\
& &f_m^{g}(\xi,k_J^2, k_f^2)=f_m^{g 0}(\xi,k_J^2, k_f^2)+\nonumber\\
& &\bar \alpha_s(\mu^2)
\int_{\xi}^1 {d\xi^{\prime}\over \xi^{\prime}} \int_0^{k_f^2/z} {dk^2\over k^2}
\left(\Delta P_{gq}(0)f_m^{S}(\xi^{\prime},k_J^2, k^2) + \Delta P_{gg}(0)
f_m^{g}(\xi^{\prime},k_J^2, k^2)\right)\Theta(k^2-k_J^2 \xi^{\prime})
\label{g},
\nonumber\\
\eqnx
}
%
%
%
%
%
%
%
%
%
%
%
%
where $z = \xi/ \xi^{\prime}$.
The inhomogeneous terms are given by the following formulae:
\eqn
f_m^{NS 0}(\xi,k_J^2, k_f^2) &=& \delta(k_f^2-k_J^2) \delta(\xi - 1)
\left({e_m^2\over \langle e^2\rangle} - 1\right)(\delta_{mq_m} + \delta_{m\bar q_m})
\label{ns0},\\
f_m^{S 0}(\xi,k_J^2, k_f^2) &=& \delta(k_f^2-k_J^2) \delta(\xi - 1)
(\delta_{mq_m} + \delta_{m\bar q_m})
\label{s0},\\
f_m^{g 0}(\xi,k_J^2, k_f^2) &=& \delta(k_f^2-k_J^2) \delta(\xi - 1)
\delta_{mg}
\label{g0}.
\eqnx
\noindent
We have solved equations (\ref{ns}), (\ref{s}) and (\ref{g}) for 
two choices of the scale $\mu^2$ which allow analytical solution of these 
equations: 
$\mu^2 = (k_J^2 + Q^2)/2$ and $\mu^2 = k_f^2/\xi$.  In the first case the 
coupling $\alpha_s(\mu^2)$ does not change along the ladder, and it is a fixed 
parameter.  In what follows we shall call this choice "the fixed coupling 
case".  Let us remind that this choice of the scale follows the convention 
usually adopted in the case of forward jet production in the unpolarised deep 
inelastic scattering \cite{FJET1}.  In the second case $\mu^2 = k_f^2/\xi$ the coupling 
changes along the ladder, and so we shall call this choice the "running 
coupling case".  One expects that the scale $\mu^2 = k_f^2/\xi$ is 
presumably too large, and that this choice will lead to an underestimate 
of the effect.  The more natural choice $\mu^2=k_f^2$ does not, however, allow 
analytical solution.  One can expect that the solution of equations 
(\ref{ns}) - (\ref{g}) with $\mu^2=k_f^2$ should lie between 
the two solutions corresponding to $\mu^2=(k_J^2 + Q^2)/2$ and 
$\mu^2 = k_f^2/\xi$.

\noindent
For the case of fixed $\bar\alpha_s$
the solutions of equations (\ref{ns}), (\ref{s}) and (\ref{g}) 
for the functions $f_m^{NS}(\xi,k_J^2, k_f^2)$ and 
$f_m^{S}(\xi,k_J^2, k_f^2)$ take the following form~:
\eqn
& &f_m^{NS}(\xi,k_J^2, k_f^2)=f_m^{NS 0}(\xi,k_J^2, k_f^2)\nonumber\\
& &+{\bar\alpha_s\lambda\over k_J^2}(\delta_{mq_m} + \delta_{m\bar q_m})
\left({e_m^2\over \langle e^2\rangle} - 1\right)
I_0[2\sqrt{\bar\alpha_s\lambda\,y(t+y)}]\Theta(k^2-k_J^2 \xi),
\label{nonsinglet}
\eqnx
%
%
%
%
\eqn
f_m^{S}(\xi,k_J^2, k_f^2) &=& f_m^{S 0}(\xi,k_J^2, k_f^2)\nonumber\\
&+&{\bar\alpha_s\lambda_0^{+}\over k_J^2}\frac{c_m^{-}}{\lambda_0^{-}-\lambda_0^{+}}
\dt_{qg}(0)\,
I_0[2\sqrt{\bar\alpha_s\lambda_0^{+}\,y(t+y)}]\Theta(k_f^2-k_J^2 \xi)\nonumber\\
&+&{\bar\alpha_s\lambda_0^{-}\over k_J^2}\frac{c_m^{+}}{\lambda_0^{-}-\lambda_0^{+}}
\dt_{qg}(0)\,
I_0[2\sqrt{\bar\alpha_s\lambda_0^{-}\,y(t+y)}]\Theta(k_f^2-k_J^2 \xi),
\label{singlet}
\eqnx
where $I_0(z)$ denotes the modified Bessel function,
\eqn
t&=&\ln\left({k_f^2\over k_J^2}\right),\\
y&=&\ln\left({1\over \xi}\right),
\lab{yt}\\
\lambda&=&\Delta P_{qq}(0),\label{lamns}
\eqnx
$\lambda_0^{+},\lambda_0^{-}$ are the eigenvalues of matrix
${\bf \dt P(0)} \equiv
\left( \begin{array}{cc} \Delta P_{qq}(0) & \Delta P_{qg}(0)\\
\Delta P_{gq}(0) & \Delta P_{gg}(0) \\ \end{array} \right )$~:
\begin{equation}
\lambda_0^{\pm} = {\Delta P_{qq}(0) +\Delta P_{gg}(0) \pm
\sqrt{(\Delta P_{qq}(0) -\Delta P_{gg}(0))^2 +
4\Delta P_{qg}(0)\Delta P_{gq}(0)}\over 2},
\label{lplus}
\end{equation}
and coefficients $c_m^+, c_m^-$ are defined as~:
\eqn
c_m^+&=&\delta_{mg}-(\delta_{mq_m} + \delta_{m\bar q_m})
\frac{\lambda_0^+-\dt P_{qq}(0)}{\dt P_{qg}(0)},\nonumber\\
c_m^-&=&-\delta_{mg}+(\delta_{mq_m} + \delta_{m\bar q_m})
\frac{\lambda_0^--\dt P_{qq}(0)}{\dt P_{qg}(0)}.\nonumber
\lab{cm}
\eqnx
This form  is typical for the double logarithmic asymptotics 
(cf. \cite{BASICS} and references therein). It has also been obtained 
for fragmentation functions \cite{WOSIEK}.

\begin{figure}[t]
    \centerline{
     \epsfig{figure=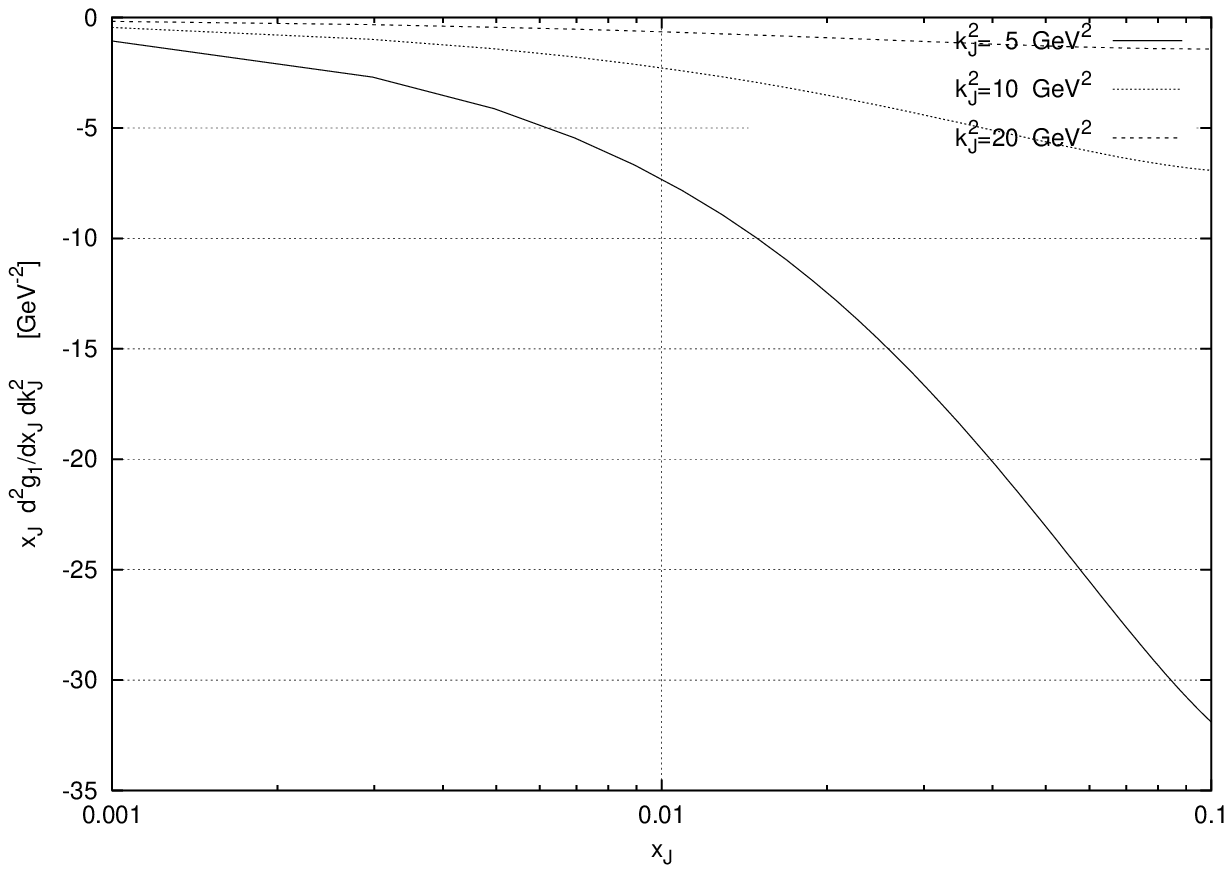,height=8cm,width=11cm}
               }
\caption{The differential spin structure function
$x_J{\partial g_1\over \partial x_J \partial k_J^2}$ for the fixed 
$\bar\alpha_s$ case plotted as the function of the longitudinal momentum fraction 
$x_J$ carried by a jet.  We show our predictions for three values 
of the transverse momentum squared $k_J^2$ of the jet i.e. for 5 GeV$^2$, 
10 GeV$^2$ and 20 GeV$^2$.  The calculations were performed  
for $Q^2=10 GeV^2$ and $x=10^{-4}$.}
\label{fig.3}
\end{figure}
%
\begin{figure}[t]
    \centerline{
     \epsfig{figure=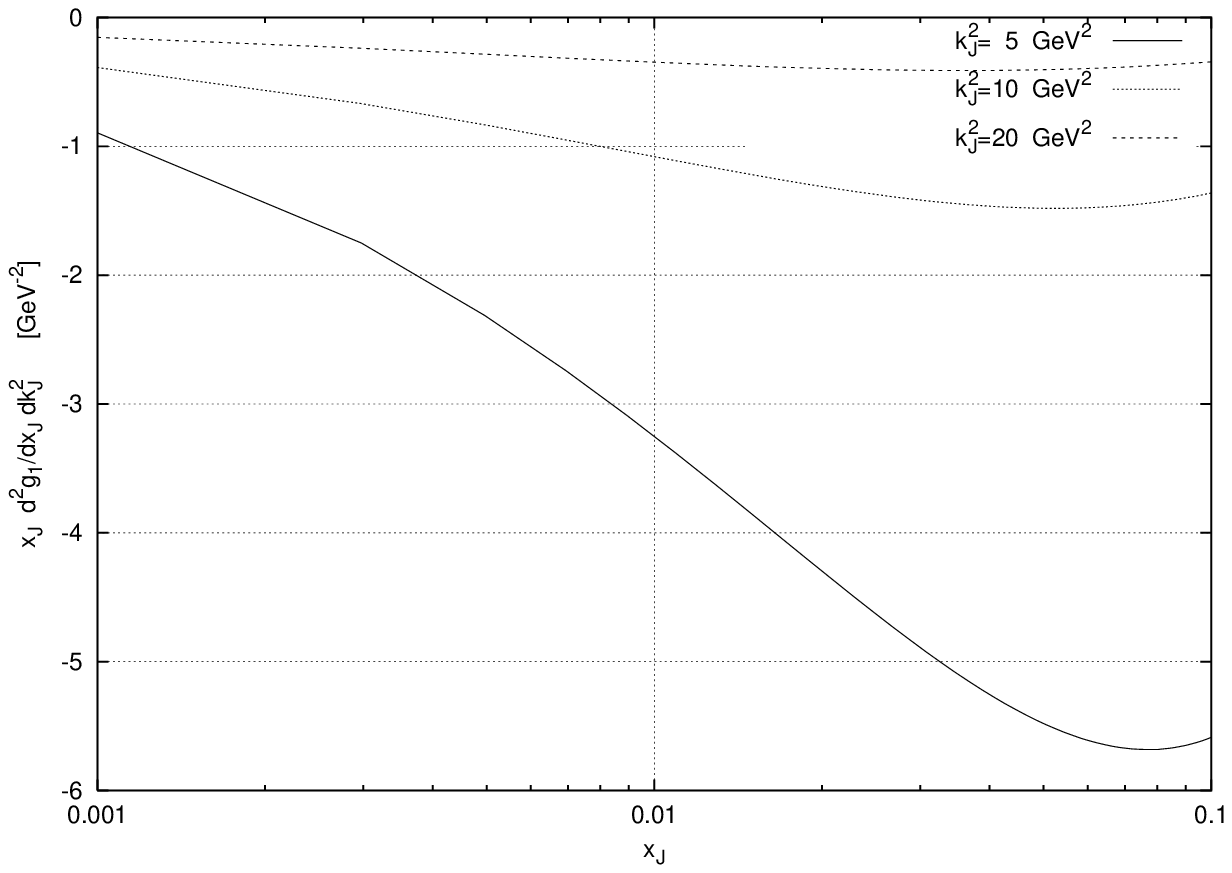,height=8cm,width=11cm}
               }
\caption{The differential spin structure function
$x_J{\partial g_1\over \partial x_J \partial k_J^2}$ for the running 
$\bar\alpha_s$ case plotted as the function of the longitudinal momentum fraction 
$x_J$ carried by a jet.  We show our predictions for three values 
of the transverse momentum squared $k_J^2$ of the jet i.e. for 5 GeV$^2$, 
10 GeV$^2$ and 20 GeV$^2$.  The  calculations 
were performed  
for $Q^2=10 GeV^2$ and $x=10^{-4}$.}
\label{fig.4}
\end{figure}
%
%
\begin{figure}[t]
    \centerline{
     \epsfig{figure=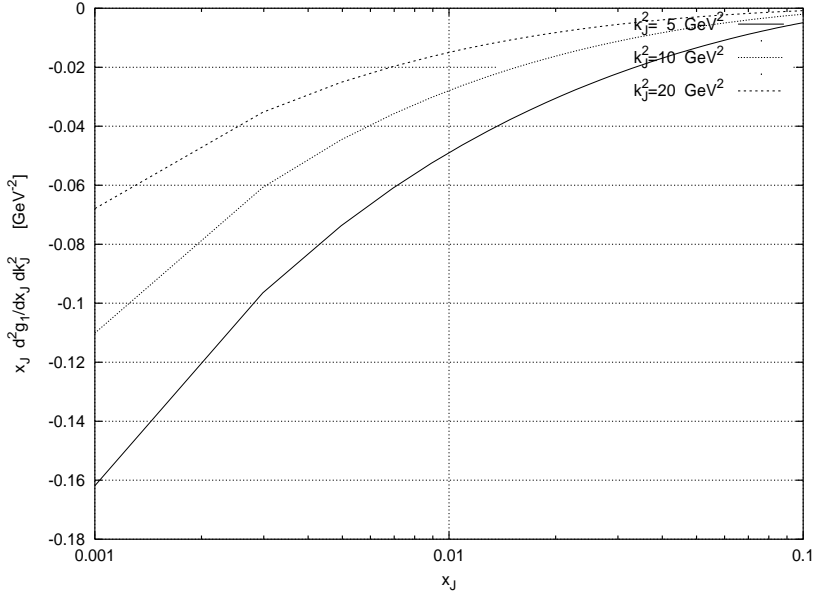,height=8cm,width=11cm}
               }
\caption{Born approximation for the differential spin structure function
$x_J{\partial g_1\over \partial x_J \partial k_J^2}$ plotted as the function 
of the longitudinal momentum fraction $x_J$ carried by a jet.  We show our predictions for three values 
of the transverse momentum squared $k_J^2$ of the jet i.e. for 5 GeV$^2$, 
10 GeV$^2$ and 20 GeV$^2$.  The  calculations were performed  
for $Q^2=10 GeV^2$ and $x=10^{-4}$.}
\label{fig.5}
\end{figure}

The solution for the case of running $\bar\alpha_s=\bar\alpha_s(k_f^2/\xi)$
for the non-singlet and the singlet components  read~:
{\footnotesize
\eqn
f_m^{NS}(\xi,k_J^2, k_f^2) &=& f_m^{NS 0}(\xi,k_J^2, k_f^2)\nonumber\\
&+&
{\bar\alpha_s(k_f^2/\xi)\lambda\over k_J^2}(\delta_{mq_m} + \delta_{m\bar q_m})
{\left({e_m^2\over \langle e^2\rangle} - 1\right)}
I_0[2\sqrt{c \lambda\,y\,\ln\frac{\rho+\rho_0}{\rho_0}}]\Theta(k^2-k_J^2 \xi),
\label{rnonsinglet}
\eqnx
\eqn
f_m^{S}(\xi,k_J^2, k_f^2) &=& f_m^{S 0}(\xi,k_J^2, k_f^2)\nonumber\\
&+&{\bar\alpha_s(k_f^2/\xi)\lambda_0^{+}\over k_J^2}
\frac{c_m^{-}}{\lambda_0^{-}-\lambda_0^{+}}\dt_{qg}(0)\,
I_0[2\sqrt{c\lambda_0^{+}\,y\,\ln\frac{\rho+\rho_0}{\rho_0}}]
\Theta(k^2-k_J^2 \xi)\nonumber\\
&+&{\bar\alpha_s(k_f^2/\xi)\lambda_0^{-}\over k_J^2}
\frac{c_m^{+}}{\lambda_0^{-}-\lambda_0^{+}}\dt_{qg}(0)\,
I_0[2\sqrt{c\lambda_0^{-}\,y\,\ln\frac{\rho+\rho_0}{\rho_0}}]
\Theta(k^2-k_J^2 \xi),
\label{rsinglet}
\eqnx
}
where~:
\eqn
\rho  &=&\ln\frac{k_f^2}{k_J^2\xi},\lab{ro}\\
\rho_0&=&\ln\frac{k_J^2}{\Lambda_{QCD}^2},
\eqnx
and
\eq
c=\frac{2}{11-2/3\,\,N_f}.
\lab{cc}
\eqx

Having calculated the unintegrated parton distributions for both the fixed
and the running $\bar\alpha_s$ case, one may express the double logarithmic
contribution $\sum_l\Phi_{m\,l}$ in (\ref{dg1}) as~:
{\small
\eqn
\sum_l\Phi_{ml}({x\over x_J},k_J^2,Q^2)=
{1\over 2} \langle e^2\rangle
\int_{x_{down}}^{x_{up}} {dx^{\prime}\over x^{\prime}}
\int_{k^2_{down}}^{k^2_{up}}
{dk_f^2\over k_f^2}
(f_m^{NS}({\bar x\over x^{\prime}},k_J^2, k_f^2)+
f_m^{S}({\bar x\over x^{\prime}},k_J^2, k_f^2)),
\label{sumphiml}
\eqnx
}
where the integration limits $x_{up}$, $x_{down}$, $k^2_{up}$ and
$k^2_{down}$ which read~:
\eqn
x_{up}=x_J,& & x_{down}=x(1+{k_J^2\over Q^2}),\nonumber\\
k^2_{up}=Q^2(\frac{x'}{x}-1),& & 
k^2_{down}=k_J^2/(\frac{x'}{x}-\frac{k_J^2}{Q^2})
\lab{limits}
\eqnx
take into account restrictions implied by the integration limits 
in (\ref{phiml}) and by the $\Theta$ functions in equations  
(\ref{nonsinglet},\ref{singlet}) etc. i.\ e.\ ~:
\eqn
k_f^2 &<&\bar W^2,\\
\bar x&<&x^{'}<x_{j},\\
k_J^2\frac{\bar x}{x^{'}}&<&k_f^2.
\lab{restrictions}
\eqnx
The latter restriction is the most important one: it follows from the ordering
of the ratio $k_n^2/x_n$ along the ladder which, in turn, determines the
infrared limit in the integration over $dk_f^2$.

We have performed numerical analysis of the low $x/x_J$ behaviour of
differential structure function ${\dd^2 g_1\over \dd x_J\dd k_J^2}$.
It was calculated by convoluting the double logarithmic
contribution $\sum_l \Phi_{m\,l}$ for the fixed and running
$\bar\alpha_s$, respectively, with the spin dependent parton distributions
$\dt p_i(x_J,k_J^2)$. The parton distributions  $\dt p_i(x_J,k_J^2)$ were obtained  
from the  input distributions at the initial scale equal 1 GeV$^2$ 
parametrised as in ref. \cite{BZIAJA,BZIAJA2} which were 
evolved using the leading order 
Altarelli-Parisi evolution for the spin dependent parton densities \cite{AP}.  
Our results are presented in Figs.\ 3 and 4
for the fixed and running $\bar\alpha_s$ case respectively.
In both cases the solutions depend strongly on the value of the
jet transverse momentum $k_J^2$ (cf. equations 
(\ref{nonsinglet}),(\ref{singlet}),(\ref{rnonsinglet}),(\ref{rsinglet})). 
One may notice that the absolute value of the differential structure function 
exhibits  very 
strong increase with the increasing value of $x_J$ (i.e. decreasing $x/x_J$). 
This increase is a direct consequence of the double $ln^2(1/x)$ resummation 
summarised in equations 
(\ref{nonsinglet}),(\ref{singlet}),(\ref{rnonsinglet}),(\ref{rsinglet}).
The absolute magnitude of the differential structure function 
which corresponds to  the running coupling case 
(i.e.  to the scale 
$\mu^2 = k_f^2/\xi$ in equations (\ref{ns}),(\ref{s}) and (\ref{g})) is significantly 
smaller than that one which corresponds to the fixed coupling , i.e. 
$\mu^2 = (k_J^2 + Q^2)/2$.  The former scale is presumably too large, and the 
latter case may be regarded as the more realistic one.

We have also calculated the differential structure function in Born 
approximation which corresponds to the replacement of functions 
$f_m^{NS}(\xi,k_J^2,k^2)$ and  $f_m^{S}(\xi,k_J^2,k^2)$ by inhomogeneous 
terms $f_m^{NS 0}(\xi,k_J^2,k^2)$ and  $f_m^{S 0}(\xi,k_J^2,k^2)$ 
defined by equations (\ref{ns0}) and (\ref{s0}). Results of this calculations 
are summarised in  Fig. 5.   We may see that the absolute  magnitude of the differential structure 
function with the effects of the double $ln^2(1/x)$ resummation (see Figs. 3 and 4) 
taken into account is significantly larger than the absolute magnitude 
of the differential  structure function  calculated in Born 
approximation.  Let us also note that the $x_J$ dependence is completely 
different in those two cases i.e. the effects of the double $ln^2(1/x)$ resummation 
generate increase of the absolute magnitude of the structure functions with increasing $x_J$ (for 
fixed $x$), while the absolute magnitude of the structure function calculated in Born approximation 
decreases with increasing $x_J$,  just following the $x_J$ dependence of 
the spin dependent parton distributions $\Delta p_i(x_j,k_J^2)$.  
Comparison with Born approximation leads to
the conclusion that the double logarithmic resummation effects should be 
clearly visible for $x_J>10^{-2}$ ($x=10^{-4}$).

To sum up,  we have estimated possible effects of the double $ln^2(1/x)$ 
resummation 
in the forward jet production in the polarised deep inelastic scattering 
at the small $x$ regime which will possibly be probed at the polarised HERA. 
We have shown that the (absolute) magnitude of the 
differential spin dependent structure function strongly increases  
with decreasing $x/x_J$,  and that it is significantly larger than the "background" 
which corresponds to Born term with the double $ln^2(1/x)$ 
resummation neglected.  Estimate of the expected rate of the forward jet 
production in polarised deep inelastic $ep$ scattering  taking into account the 
HERA acceptance is in progress.      

\section*{Acknowledgments}

This research has been supported in part by the Polish Committee for Scientific
Research grants 2 P03B 184 10, 2 P03B 89 13 and 2P03B 04214  and by the
EU Fourth Framework Programme 'Training and Mobility of Researchers', Network
'Quantum Chromodynamics and the Deep Structure of Elementary Particles',
contract FMRX--CT98--0194.

\end{document}